## ARTICLE

[1]*Biological and Environmental Engineering Department, Cornell University, Ithaca, NY, USA*

[2] *Meinig School of Biomedical Engineering, Cornell University, Ithaca, NY. USA.*

[3]*Sibley School of Mechanical and Aerospace Engineering, Cornell University, Ithaca, NY. USA.*

[4]*Department of Pathology, Albert Einstein College of Medicine, 1300 Morris Park Avenue, Bronx, NY, USA 10461*

# Viscoelastic properties of tumor spheroids revealed by a microfluidic compression device and a modified power law model

Mrinal Pandey[1], Bangguo Zhu[3], Kaitlyn Roach[2], Young Joon Suh[1], Jeffrey E Segall[4], Chung-Yuen Hui[3], and Mingming Wu[1]

## Abstract

Clinically, palpation is one of the important diagnostic methods to assess tumor malignancy. In laboratory research, it is well accepted that the bulk stiffness of the tumor and the surrounding tissue is closely correlated with the malignant state of the tumor. Here, we postulate that, in addition to tumor stiffness, tumor viscoelasticity - the fact that tumor tissue takes time to bounce back after compression, can also be used to evaluate the tumor malignancy state. In this work, we characterized the viscoelastic properties of breast tumor spheroids using a recently developed microfluidic compression device and a theoretical power law model. Breast tumor cells at varying malignant levels; a non-tumorigenic epithelial (MCF10A), moderately malignant tumor (MCF7) and triple negative metastatic tumor (MDA-MB-231) cells were used. Spheroids embedded within a 3D extracellular matrix were periodically compressed, and their strain responses were recorded using microscopic imaging. Our results revealed that the measured strain relaxation curves can be successfully described by a modified power law model, demonstrated that non-tumorigenic tumor spheroids were more elastic, exhibited shorter relaxation time and less plasticity than those of tumorigenic spheroids. Together, these results highlight that viscoelastic properties in addition to bulk stiffness of the tumor spheroids can serve as a complementary mechanical biomarker of tumor malignancy and demonstrate the validity of a modified power law model for the mechanical characterization of a living tissue.

## Introduction

Tumor mechanics emerges as a key biomarker of tumor malignancy state, with extensive studies demonstrating that increased tissue stiffness promotes tumor progression [1-3]. This elevated stiffness arises from the uncontrolled proliferation of tumor cells [4] and the stiffening of the surrounding extracellular matrix [5, 6]. Paradoxically, multiple studies have shown that, at the single cell level, malignant cells are often softer than their non-malignant counterpart [7, 8]. We know that tumors, or living tissues in general, are inherently complex and highly adaptive, capable of dynamically remodelling the mechanical properties of their surrounding matrices in response to changes in the microenvironment [9, 10]. It is thus important to develop strategies to investigate tumor mechanics using physiologically realistic experimental conditions and at multiple time and length scales.

Tumor mechanics studies today have largely focused on the bulk stiffness of the tumor cell/tissue or surrounding ECMs. It is only recently that the viscoelastic properties of ECM instead of stiffness alone have been found to play important roles in cell mechano-transduction [11-13] and tumor invasion[14-17]. This is not a surprise since tumor invasion is a dynamic process[9]. In a landmark study, Chaudhuri et al [16] highlighted the importance of examining viscoelastic properties of living materials, emphasizing their critical roles in regulating cell migration behavior. In contrast, to date, very little is learned about the viscoelasticity of tumor tissue, and how they correlate with tumor malignant state[18].

Examining the viscoelastic properties of tumors is also crucial from a basic science viewpoint. We ask the basic question: how can we theoretically model the mechanics of living materials? The simplest class of solid materials is the Hookean solid where the stress is proportional to the strain. In contrast, biological materials are complex in their behavior; they often deform and adapt when subject to external stresses. Type I collagen, a primary protein that provides architectural support to mammalian cells, is an example [16, 19]. Collagen consists of a cross-linked fibre network that aligns when stretched and buckles when compressed. This leads to the complex nonlinear elastic and anisotropic material properties of collagen [20, 21]. Living materials such as tissues (e. g. cell embedded collagen) enhance the complexity of the materials further. Here, the basic constituents, cells, can sense, move, as well as modify the extracellular matrix (ECM) architecture. For example, tumor cells are mechano-sensors; they move and re-arrange the surrounding ECM spatially when a tissue is subjected to an external force [22]. In this context, it adds one more dimension to the material mechanics.

Commercially available parallel rheometers are the current workhorse for characterizing the mechanics of biological materials today. In parallel rheometers, biological samples are sheared or compressed between two parallel plates, and a stress–strain relationship is obtained. While straightforward, this setup is not easily compatible with living tissues or optical imaging, limiting its use for living materials. AFM addressed some of these issues by enabling force measurements in 3D environments [23-25], and optical compatibility but its fine tip restricts analysis to localized single-cell mechanics. Micro-pipette aspiration [26-28] has been used to contribute to valuable insights of tissue mechanics although the data analysis and experimentation are challenging. To overcome these limitations, microfluidic platforms have emerged as promising tools for studying cell and tissue mechanics in biologically relevant settings, with better visualization and fine control over bulk cell and tissue mechanics. Examples include PDMS micro-piston systems [29-31], and constriction based devices that probe mechanical resistance by forcing aggregates through narrow channels [14]. Despite their innovations, each of these methods has inherent limitations in terms of throughput, physiological relevance, or spatial resolution and optical compatibility. To address these gaps, in this manuscript, we have developed an integrated approach that combines a microfluidic compression device that provides a physiologically realistic 3D environment to tissue with a modified power-law model to investigate the viscoelastic mechanical properties of tumor spheroids.

## Results and Discussion

### Microfluidic compression device setup and calibration

The key capability of the microfluidic device is to provide a well-controlled compression to the tumor spheroids embedded in a physiologically realistic 3D ECM and compatible with microscopic imaging (Fig. 1). The device consists of 12 functional units, of which 6 are control units and the other 6 are compression units (Fig. 1A). The cross-sectional view of each unit is shown in Fig. 1B. The main features of the compression unit are the cell chamber layer (L1) that holds the spheroid embedded ECMs and the pressure control unit (L2+L3) that provides well defined compression to the spheroid. To calibrate piston displacement under varying pressure levels, we measured piston displacement by locating the focal point of 1 μm green, fluorescent beads on the piston surface as a function of the pressure level (Fig. 1C). Briefly, beads were placed at the bottom of the piston and upon applying pressure, piston displacement caused beads to shift vertically and defocus. The displacements were measured by refocusing the beads using a motorized stage [32]. The piston displacement versus pressure curve was validated against the COMSOL computation as shown in Fig. 1C, where the red line is the result of COMSOL simulation and dots are experimental measurements. The actual device geometry (Fig. S1A) is chosen for COMSOL simulation. The simulation was performed using an axisymmetric model and the PDMS was modelled as a linear elastic material with modulus of 1.33 MPa and Poisson's ratio of 0.49. In the simulation, the bottom glass was modelled as a rigid material with fixed constraints, the sides of the device were fixed, and the pressure was applied to the top surface Fig.S1B. An extra fine mesh was applied across the model to ensure numerical accuracy.

# ARTICLE

For a typical experiment, the cell chamber layer (L1) was first placed on a standard glass slide (75 mm × 25 mm x 1mm) facilitating microscopic imaging of the embedded tumor spheroids. Each cell chamber is 3 mm in diameter, and 500μm in depth. The rim of each cell chamber contains 4 slits, each 0.5 mm wide, to allow media transport during compression. For more details on device design and development please refer to [33]. Second, tumor spheroids were embedded in a 1.5 mg/ml type I collagen at a concentration of 1296 spheroids/mL and were introduced into the cell chamber. Note that on average 2-3 collagen spheroids were embedded per cell chamber to avoid potential spheroid-spheroid interaction during the experimental run. Also, a 1.5 mg/ml collagen matrix was used to provide a soft matrix that minimally interferes with the measurement of the mechanical properties of the spheroids, while ensuring sufficient anchorage to prevent bulk movement during cyclic compression. Notably, our lab's previous studies using the same microfluidic device to calculate spheroid modulus reported comparable relaxation time responses even when spheroids were suspended in media alone[33]. Third, the pressure control unit (L2+L3) was carefully aligned and placed atop L1, ensuring that each piston was cantered precisely over the corresponding cell chamber. This alignment is important for uniform compressive forces across each individual spheroid. Lastly, the mid-z plane of the spheroid is imaged using a bright field microscope (Fig. 1D). A convolutional neural network, U-Net [34] segmentation algorithm, was used to trace the spheroid boundaries. The red/green outlines in Fig. 1D are the traced spheroid outlines with/without compression respectively. The outline of the spheroid is used to compute the spheroid area $A$ in the mid-z plane, and the spheroid diameter is computed using $D = \sqrt{4A/\pi}$. We see that the spheroid increases in its diameter upon compression (Fig. 1E).

Uniform spheroids were generated using a standardized agarose microwell array technique previously developed in our labs [35]. Fig. 1F shows that there are no significant differences in initial spheroid sizes among the different cell lines, ensuring a fair basis for comparative mechanical testing. Here, tumor spheroids were made with tumor cells of increasing malignancy states: a triple-negative metastatic basal breast cancer cell line (MDA-MB-231), a moderately malignant luminal breast cancer cell line (MCF7), and a non-tumorigenic breast epithelial cell line (MCF10A).

**Spheroid size response under cyclic compression exhibits viscoelastic material behavior**

To study the viscoelastic properties of tumor spheroids, we investigated the tumor spheroid size response under a controlled dynamic compression of the spheroids (Fig. 2). A square pressure wave with a max pressure of 14kPa and a period of 40 seconds is applied to the spheroid through the pressure control unit (Fig. 2A, and SMovie1 and SMovie2). The spheroid is imaged at mid-z plane optically, and its size responses are shown in Fig. 2B. Using the spheroid images, we calculated the normalized area change of the spheroid, $(A - A_0)/A_0$, where A is the spheroid area in the mid-z plane after compression, and $A_0$ is the initial spheroid area before compression. Looking closely at the spheroid area response curves, we notice that upon compression, it takes a few seconds to reach the near steady state; upon release of the compression, the area change undergoes a jump first and then relaxes to a lower near steady state (Fig. 2B). Both cell types show an immediate deformation (elastic response), followed by a gradual increase in deformation over time (viscous response).

We computed strain response curves using the data in Fig. 2B as shown in figure 2C. We converted the normalized area change into height change (Δh/h) using a geometric relationship assuming that the spheroid is incompressible at the time scale of our interest and a neo-Hookean material for the spheroid. The detailed computation is in supplementary theoretical model section 3 and Fig. S4 and S5. Δh/h here is defined as the linear spheroid strain in our measurements. To illustrate the concept of a strain response of a elastic material, a solid brown line for elastic material is added to Fig. 2C as reference.

To evaluate the viscoelasticity of the tumor spheroid, we focus on the relaxation phase of the strain response curve. An important feature of a linear viscoelastic material with a single time scale is that the strain response to a sudden release of a constant compression decays exponentially[36, 37]. Here, we found that the relaxation response of tumor spheroids exhibited a clear power-law relationship when plotted on a log-log scale. This indicated that multi-time scales are involved in the response curves. Fig 2D-E show the log-log plot relationship of the single malignant and non-malignant spheroids which are shown in Fig 2B. To estimate the time, it takes for spheroids to bounce back after compression, we analysed the absolute slope values of the relaxation curves across all spheroid types and found that the non-tumorigenic MCF10A spheroids exhibited the fastest recovery (see Fig. 2F) while the metastatic MDA-MB-231 spheroids the slowest, and MCF7 spheroids in the middle. Interestingly, while the recovery slopes of MCF7 and MCF10A spheroids were not significantly different, the slopes of MDA-MB-231 and MCF10A spheroids differed significantly. The faster recovery of MCF10A spheroids indicates an overall more elastic behavior, whereas the slower recovery of malignant MDA-MB-231 spheroids is indicative of a more fluid like behavior.

Overall, the differential strain relaxation curves for MDA-MB-231 spheroids versus MCF10A spheroids may come from the differential cell-cell adhesion of these two types of spheroids. In MCF10A spheroids, cells adhere to each other via a direct cell-cell adhesion molecule, E-cadherin. While the MDA-MB-231 cells adhere to each other through a secondary adhesion, cell-ECM, and ECM-cell adhesion via β1-integrin[38, 39]. The tightly cohesive cell-cell adhesion in MCF10A spheroids may lead to a more solid like behavior than the MDA-MB-231 cell spheroids where cell-cell adhesion is weak. This is consistent with the fast bounce back time for MCF10A spheroids. In a separate note, single cell stiffness of MCF10A is stiffer than MDA-MB-231 cells, this could also lead to a high elasticity which may contribute to the faster rebound time seen in Fig. 2F.

Importantly, all spheroid types, regardless of malignancy status, exhibited power-law relaxation behavior, reinforcing that living biological materials display complex rheological behavior[40] rather than simple linear viscoelasticity. This characteristic may reflect dynamic cytoskeletal remodeling, cell-ECM interactions, and fluid redistribution occurring over a broad range of timescales. More studies matching the timescales (in minutes to hours) of these cellular phenomena can give deeper insight into cellular mechanics and invasion characteristics.

**Viscoelastic properties of tumor spheroids revealed by a modified power law model**

The power law behavior observed in the spheroid size relaxation curves (Fig. 2D, E) suggest that the material model for tumor spheroids requires a multi-time scale model beyond the simple Maxwell or Kelvin-Voigt models. Here, we propose a modified power law model that has been successful in describing viscoelastic properties of soft materials previously[41, 42]. This model, originally described in [43] has a stress relaxation function of:

$$Y(t) = E_\infty + \frac{E_0 - E_\infty}{\left(1 + \dfrac{t}{t_R}\right)^m} \qquad (1)$$

where $E_0 = Y(t=0)$ is the instantaneous modulus, $E_\infty = Y(t \to +\infty)$ is the relaxation modulus, $t_R$ and $m$ are two material fitting parameters characterizing the relaxation speed. To validate that this model applies to our system, we first obtain the analytical function of the strain relaxation curve upon the release of the compression similar to our experimental setting (See Supplementary theoretical models' section 1. and Equations S10a, b) and ref [44]. We then fit the experimental strain curves to the analytical function (S10a,b) which results in fitted parameters m and $t_R$ (see Fig S2). Fig. 3A shows an excellent fit with two adjustable parameters $t_R$ and $m$ which shows the applicability of the modified power law model (Eqn 1) to our data.

To characterize the stress response of the tissue, we computed stress relaxation curves using the fitted parameter m and $t_R$ and Eqn 1 for the three spheroid types (Fig. 3B). Here, the half time for stress to relax to its final value is computed. We find that the non-tumorigenic tumor spheroid (MCF10A) relaxes much faster than the malignant MDA-MB-231 spheroids, an average value of $1.27 \pm 0.57$ s versus $5.57 \pm 1.34$ s (Figure 3C). Our results reveal clear differences between malignant and non-malignant spheroids in its half time, which is consistent with our estimate about the bouncing back time in strain (Fig. 2F). This provides a foundation to distinguish differential malignant states of the spheroids based on viscoelastic properties rather than relying solely on elastic measurements alone.

To further quantify the frequency-dependent dissipation, we calculated the normalized effective viscosity $\eta(\omega)/E_0$, derived from the imaginary component of the complex modulus associated with our stress relaxation function and divided by the instantaneous modulus (see Theoretical models' section 2 and ref [45]). Interestingly, the effective viscosity profiles of the malignant (MDA-MB-231) and non-tumorigenic (MCF10A) spheroids revealed a frequency dependent behavior and a crossover at log frequency of -1 corresponding to period of ~10s (Fig. 3D). At lower frequencies within our computed window (log f ≈ -1 to -3.8, corresponding to timescales of ~10s -100 min), MDA-MB-231 spheroids exhibited higher effective viscosity than MCF10A, indicating a more viscous-like mechanical behavior in malignant spheroids. This trend reversed at higher frequencies (log f ≈ -0.9–1.2, corresponding to timescales of ~8s to -0.06 s), where malignant spheroids appeared less viscous than their non-tumorigenic counterparts. The observed frequency dependence underscores that the apparent mechanical properties of tumors depend on probing frequency. Thus, a comprehensive evaluation of tumor mechanics requires measurements across a broader frequency spectrum. In addition to these two cell lines, we also looked at MCF 7 cells and consistent to half relaxation data, effective viscosities of MCF 7 were in between MDA-MB-231 and MCF10A (see Fig. S3).

**Plasticity and spheroid expansion rate under long time compression**

Plasticity, a permanent deformation when subjected to an external force, is an important characteristic of biological materials. In tumors, tissue plasticity is often correlated with tumor malignancy progression and invasion [46-48]. To quantify plasticity, we monitored spheroid strain relaxation under prolonged periodic compressions using a 40-min square pressure wave (see SMovie3 and SMovie4). Across all three cell lines, spheroid strain increased continuously during compression, and upon release, the strain did not return to its original value (Figure 4A1, B1, C). To quantify this, we measured the residual strain by calculating the difference of the strain at the end of two successive cycles (See Fig, 4A1). Residual strain provides a measure of irreversible deformation accumulated within the spheroids after each cycle of compression. We found there to be a significant difference in residual strain between malignant and non-malignant cell lines. The malignant MDA-MB-231 spheroid has an average residual strain of $0.166 \pm 0.038$, and in contrast, the non-tumorigenic MCF10A spheroid has an average residual strain of $0.038 \pm 0.011$. The elevated residual strain in MDA-MB-231 spheroids reflects a higher degree of plasticity and potentially structural remodelling in comparison to non-malignant spheroids. In addition, we converted this area deformation to height deformation using FEM analysis discussed in Supplementary theoretical model section 3 and quantified plasticity using the linear strain values (see Fig S6).

We observed a continuous increase in spheroid area during compression phase as seen in Fig. 4 A1, B1 for both types of spheroids. The gradual increase of spheroid area under compression signifies an expansion of the spheroid into the surrounding matrix. We quantified this expansion using a linear fit to the spheroid area change as a function of time (See Fig. 4A2, B2) and Figure S7. For all spheroids, we observed a near linear expansion in area with time. The linear fit to the data showed that the metastatic MDA-MB-231 spheroids exhibited the steepest slopes with an average value of $(7.41 \pm 1.34) \times 10^{-5}$ s$^{-1}$. In contrast, MCF10A spheroids displayed significantly lower slopes, with an average value of $(1.98 \pm 0.534) \times 10^{-5}$ s$^{-1}$, see (Fig. 4D). This is consistent with the understanding that normal epithelial cell line MCF10A has more stable cytoskeleton and stronger cell-cell adhesion that leads to slower expansion than malignant MDA-MB-231[49, 50].

Together, these findings revealed that tumor spheroids exhibit plasticity under periodic compression, with malignant spheroids demonstrating enhanced plasticity than the non-tumorigenic spheroids. These results complement our short-timescale observations and underscore the importance of investigating tumor mechanics across multiple timescales to fully capture the dynamics of living materials.

## Conclusion and future perspectives

In this work, we investigated the viscoelastic properties of breast tumor spheroids using an integrated microfluidic compression device and theoretical modelling. Our results showed that breast tumor spheroids at three different malignancy states followed a modified power law model. Notably, their mechanical phenotypes varied significantly with malignancy state: upon release of a constant compressive strain, non-tumorigenic spheroids recovered rapidly and behaved more elastically on a short time scale (seconds) and possessed less of a permanent deformation (plasticity) than those of malignant tumors at longer time scale (minutes). These observations are consistent with previous work where stable cytoskeleton and stronger e-cadherin mediated cell-cell adhesion has been correlated with less invasion[49, 50].

The modified power law model used here provided a simple and generic method for modelling living materials where multiple time scales are involved. In the future, we plan to interrogate biological activities at various time scales individually including cytoskeletal molecule dynamics, ECM remodelling and cell migration to understand the nature of multi-time scale dynamics presented here. We also hope that the modified power law model can be used for characterizing mechanics of other tissue types to further understand the biological implications of viscoelastic properties of living tissue.

## Materials and Methods

### Cells, spheroids, and 3D spheroid culture preparation.

**Cells:** In this study, three different breast tumor cell lines were used. They are MCF10A, non-tumorigenic epithelial cell line; MCF7, moderately malignant, estrogen receptor positive (ER +) cell line, and MDA-MB-231, metastatic breast cancer adenocarcinoma cell line[51]. Growth medium for MCF10A cells was composed of DMEM/F-12 medium (Cat. 11320033, Gibco), 5% donor horse serum (Cat. S12150, Atlanta Biologicals), 20 ng/mL human EGF (Cat. PHG0311, Gibco), 0.5µg/mL hydrocortisone (Cat. H0888–1G, SigmaAldrich, St. Louis, MO), 100 ng/mL Cholera Toxin (resuspend at 1 mg/ml in sterile DI H2O, Cat. C8052-.5MG, Sigma-Aldrich), 10µg/mL insulin (Cat. 10516-5ML, Sigma-Aldrich), and 1% antibiotics (Gibco). The growth medium for MCF7 cells was composed of MEM Alpha (Cat.12571-063, Gibco), 10% fetal bovine serum (Cat. S11150, Atlanta Biologicals, Lawrenceville, GA), and 1% antibiotics (100 units/mL penicillin and 100µg/mL streptomycin, Cat. 15140122, Gibco). Growth medium for MDA-MB-231 cells was composed of DMEM high glucose medium (Catalog No. [Cat.] 11965092, Gibco, Life Technologies Corporation, Grand Island, NY), 10% fetal bovine serum (Cat. S11150, Atlanta Biologicals, Lawrenceville, GA), and 1% antibiotics (100 units/mL penicillin and 100µg/mL streptomycin, Cat. 15140122, Gibco). All cell lines were cultured for up to 20 passages and used at 70-90% confluency. Spheroids of all three cell lines were prepared using DMEM/F12 media.

**Spheroids:** Uniform spheroid size was generated using an agarose microwell array technique previously developed in our labs [35, 52]. Briefly, a silicon master 36 × 36 microwell array platform was first fabricated in the Cornell Nanofabrication facility using a one-layer photo-lithography method. Microwells from this wafer were patterned using soft lithography on a 1 mm thick and 1x1cm agarose gel, which promoted clustering of cells and spheroid formation due to its low cell adhesion. Microwells for MDA-MB-231 and MCF10A spheroids were 350 µm in diameter and depth, while the microwells for MCF7 spheroids had a diameter of 200 µm and depth of 230 µm. Microwell dimensions, as well as cell seeding density, together determine the size of spheroids. We placed one agarose microwell each in 6 different wells of a 12-well plate (Cat. #: 07-200-82, Corning). In each well, 3 million cells were suspended in 2.5 ml of DMEM/F12 growth medium. The plate was gently placed in a 5% CO2 incubator at 100% humidity for seven days, with media replenished on day 3 and day 5. We filtered using a Falcon Cell Strainer (Cat. #: 352360, Corning) with 100 µm pores to ensure the uniformity of the spheroid size. The architecture of each spheroid type was different: MCF7 and MCF10A spheroids were more compact and were formed overnight, whereas MDA-MB-231 spheroids were less compact and took 4 to 7 days to form uniformly sized spheroids. It is important to note that rich media and 7 days culture are important to form uniform sized MDA-MB-231 tumor spheroids. For consistency and cross comparison all spheroids were harvested on day 7 for all cell lines, and one array of microwell was used for each experiment.

**Spheroid embedded ECM:** To make 3D tumor spheroid cultures, we suspended spheroids in a 1.5 mg/ml type I collagen matrix (rat tail tendon Cat. #: 354249, Corning). For each experiment, 200 µl of spheroid embedded collagen mixture was prepared with a collagen concentration of 1.5 mg/ml. To do this, 27.1 µl of collagen stock (11.07 mg/ml) was first titrated with 0.6 µl 1 N NaOH and 20 µl 10X M199 (Cat. #: M0650-100MI, Sigma) to yield a final pH of ~7.4. Then, 152.3 µl of spheroids with DMEM, MEM α or DMEM/F12 GM for MDA-MB-231, MCF7 and MCF10A spheroids respectively were added to reach a final volume of 200 µl.

**Microfluid compression device fabrication**

**Soft Lithography:** The microfluidic device consists of three layers, all of which were made of PDMS and molded from Silicon master molds. For the making of the silicon master mold, see reference[33]. All three layers of the microdevice were fabricated using soft lithography techniques using the silicon masters. The sample chamber layer L1 is made via a PDMS double casting method. Briefly, first we make a PDMS master mold using 25 g of 10:1 polydimethylsiloxane (PDMS) mixture, degassed, and poured into a 5.3 mm thick polycarbonate frame placed around the feature on the silicon wafer. A sheet of laser printer transparency film was placed on top to create a smooth surface, followed by a glass panel with a weight to ensure even thickness. The PDMS was then cured at 60°C overnight. The cured PDMS was cut, removed from the silicon master, plasma cleaned for 1 minute, and treated with in-house FOTS setup overnight. This PDMS mold was then used for second casting to get actual L1 on a glass slide using 10g of 10:1. A clean 25 mm x 75 mm x 1 mm glass slide was taped on all four sides onto the bottom of a petri dish, and PDMS was poured on top of the glass slide and the feature side of the PDMS mold. After removing all bubbles in a vacuum chamber, the PDMS mold was flipped onto the slide, secured with tape, weighted down flush such that the height of the device is defined by the PDMS mold. The assembly was cured overnight at 60°C and then carefully peeled away, leaving the L1 securely attached to the glass slide.

To fabricate the piston membrane layer L2, 10 g of 10:1 PDMS was prepared and degassed. A 340 µm thick rectangular frame was placed around the feature on the Silicon wafer, and PDMS was poured into the frame. A sheet of transparency film was carefully placed on top without introducing bubbles, followed by a glass slide and a weight. The PDMS was cured at 60°C overnight. After curing, the frame and excess PDMS were carefully removed, leaving only the center PDMS piece on the wafer ready to be bonded with the L3 later.

To fabricate the pressure chamber layer L3, 25 g of 10:1 PDMS was prepared and degassed. A 5.3 mm thick rectangular frame was positioned on the Silicon wafer, and PDMS was poured into it. A sheet of transparency film was placed on top without introducing bubbles, followed by a glass panel and a weight. The PDMS was cured at 60°C overnight. After curing, the PDMS was cut and removed from the wafer, and a 1.5 mm hole was punched at the inlet using a biopsy punch for pressure delivery.

**Microfluidic compression device assembly and operation**

The pressure control unit is made by bonding L2 and L3. Both L2 (attached to the wafer) and L3 PDMS were plasma cleaned for 1 minute in a plasma cleaner (PDC-001-HP, Harrick Plasma, Ithaca, NY, USA). The features were then carefully aligned under a microscope and bonded. The bonded layers were placed in a 90°C oven for 15 minutes to strengthen the bond, then cooled slowly to room temperature. Finally, the PDMS was carefully detached from the wafer and cut to match the size of L2. Then, 4 access ports were made using 2 mm biopsy punches to allow media replenishment and gas exchange.

Prior to experiment, the pressure control chambers were filled with water, and the surface of the sample loading chamber was activated for better adhesion with type 1 collagen gel. For the pressure control unit preparation, the pressure control unit was plasma cleaned for 30 seconds and submerged under water overnight to fill the pressure control chamber with water. Then, a tubing (OD and ID of 1/16" and 1/32", respectively) pre-filled with water from the pressure controller (Elveflow OB1 MK4, Paris, France) was connected to the 1.5 mm inlet on the device. To prepare the sample loading chamber layer L1, the well surface was treated such that collagen matrices can be bonded to the surface. Here, each well in the L1 was selectively plasma cleaned for 20 seconds with a mask, treated with 1% Polyethylenimine (PEI) for 10 minutes, washed with PBS once, and treated with 0.5% glutaraldehyde for 30 minutes. After three washes with PBS, the L1 is ready for use in the experiment.

On the day of the experiment, 8 µl of 1.5 mg/ml collagen embedded tumor spheroids were loaded to the sample chamber L1. The pressure control unit was then aligned under a microscope and lowered onto the L1. The three layers were then sandwiched between a metal frame and a Plexiglas window, secured with screws to ensure uniform pressure around the entire assembly as shown in Fig. 1A. The assembled device was then placed in a 37°C and 5% $CO_2$ incubator for 45 minutes for collagen polymerization. After the polymerization, the device was carefully filled with cell media through the four access ports. Lastly, the device was connected to the Elveflow pressure controller and placed on a temperature, humidity, and $CO_2$ controlled microscope stage for imaging.

The pressure in the pressure control unit was regulated using a commercially available pressure controller (OB1 MK4, Elveflow Inc., Paris, France), operated via Elvesys software. The software allows users to generate pressure waveforms with customizable profiles. The piston's stability and responsiveness under pressure inputs were validated, for details refer[33]. An external pressure source (Model 3 Compressor, Jun-Air, MI, USA) supplied pressure to the OB1 MK4 controller.

**Imaging and data analysis**

All images were taken using an inverted epi-fluorescent microscope (IX81, Olympus America, Center Valley, PA, USA) with a CCD camera (ORCA-R2, Hamamatsu Photonics, Bridgewater, NJ, USA). In all the experiments the middle z-plane of the spheroids was captured using a 20X objective (Olympus, NA = 1) in bright field. The scope has a stage incubator (Precision Plastics Inc., Beltsville, MD, USA) that maintained a temperature of 37 °C, humidity of ~70%, and 5 % $CO_2$ level. The setup was placed on the automated X-Y microscope stage (MS-2000, Applied Scientific Instrumentation, Eugene, OR), and images were taken at ~ 16 frames per second for 40s square wave spheroid mechanics data and at 1 frame per minute rate for 40-minute tumor mechanics data using CellSens software (Olympus America, Center Valley, PA, USA). The 16 frames per second frame rate allows us to observe the compression and relaxation dynamics of the spheroid on a timescale of ~60 ms.

## Figures:

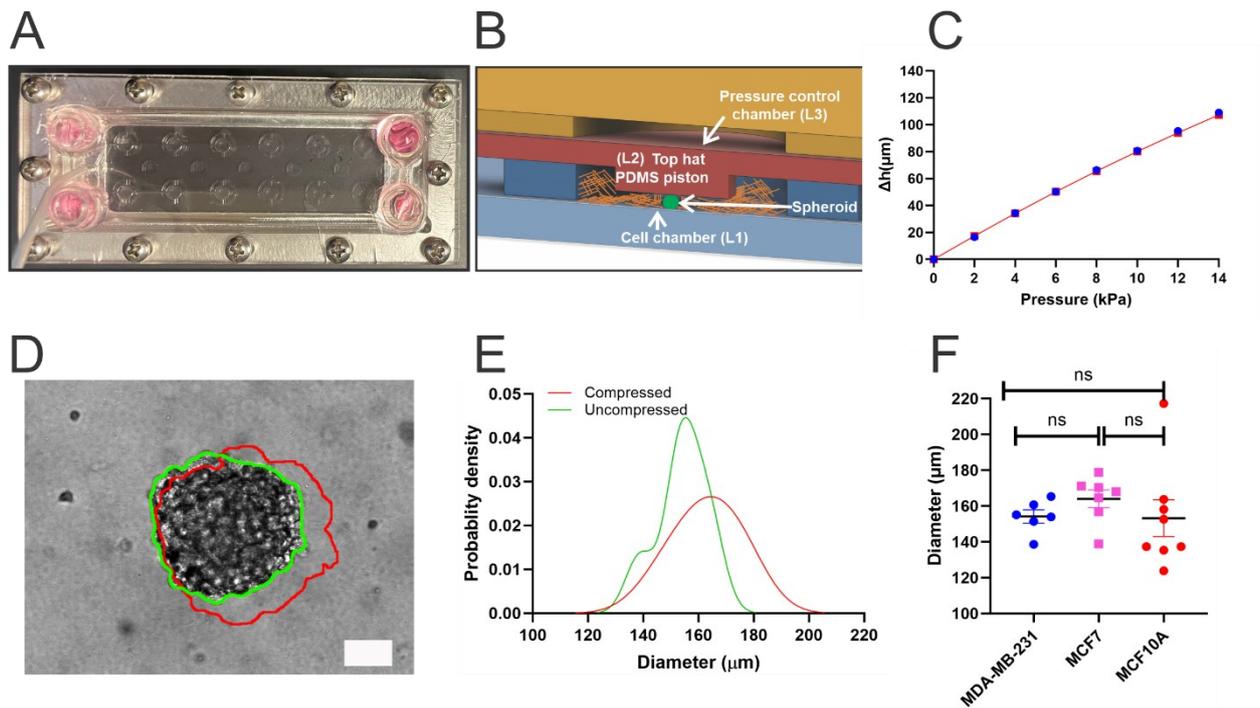

**Figure 1: Experimental setup and device calibration. (A)** Top view of a fully assembled microfluidic device. The device has a total of 12 functional units and is sandwiched between a stainless-steel frame and a polycarbonate manifold. **(B)** A schematic drawing showing the cross section of one compression unit**.** Each unit consists of three PDMS layers. The first is a cell chamber layer (L1), where the spheroid embedded collagen is placed. The second is a deformable PDMS membrane with a top hat shape piston (L2) for pushing onto the tumor spheroids with a flat surface. The third is a pressure control chamber layer (L3) that are connected to an external pressure control device. L2 and L3 are bonded to form a pressure control unit to apply a desired pressure onto the spheroids. **(C)** Device calibration. Deflection of the piston as a function of pressure applied (dots are experiments and the red line is COMSOL computation result). **(D)** A bright field image of a MDA-MB-231 spheroid embedded in collagen. The red or green outline marks the peripheral of the spheroid with and without compression respectively. The compression pressure is 14kPa. The scale bar is 50 μm. **(E)** Size distribution of MDA-MB-231 spheroids before and after compression. **(F)** Measured diameter of spheroids made of three different breast cell lines with increasing malignancy before compression; non-tumorigenic MCF10A, ER positive tumorigenic MCF7 and triple negative metastatic MDA-MB-231.

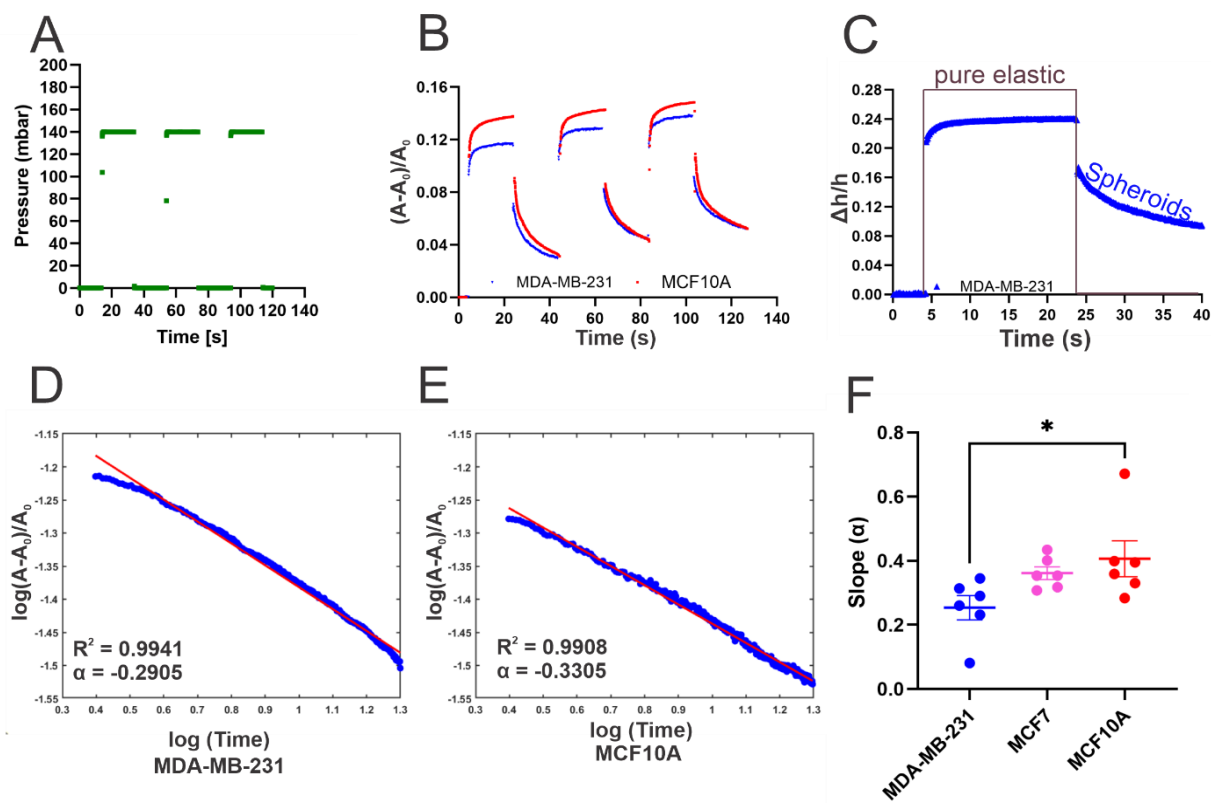

**Figure 2: The spheroid size response curve follows a power law (A)** A square pressure wave applied to the pressure control unit, with a maximum pressure of 14 kPa and a period of 40 s. **(B)** Normalized tumor spheroid area changes under the cyclic compression of the square pressure wave shown for a malignant and a non-malignant tumor spheroid. Here, the diameter of the MDA-MB-231 spheroid is 154.20 ± 4.07 μm, MCF10A is 143.14 ± 6.2 μm. **(C)** Computed strain response of a malignant (MDA-MB-231)) tumor spheroid using data in (B). For reference, a brown line is added to show the strain response of a pure elastic material. The figure is adapted from [53] **(D-E)** Normalized spheroid area change response curve of MDA-MB-231 (D), and MCF10A (E) spheroids. Dots are experiments and lines are fits to linear function. α is the fitted slope and $R^2$ is the goodness of fit, where 1 indicates a perfect fit. **(F)** Slopes of the strain response curve of three different cell lines. The stars were obtained using a nonparametric t-test (Mann–Whitney test with ****: $P < 0.0001$, ***: $P < 0.001$, **: $P < 0.01$ and *: $P < 0.05$).

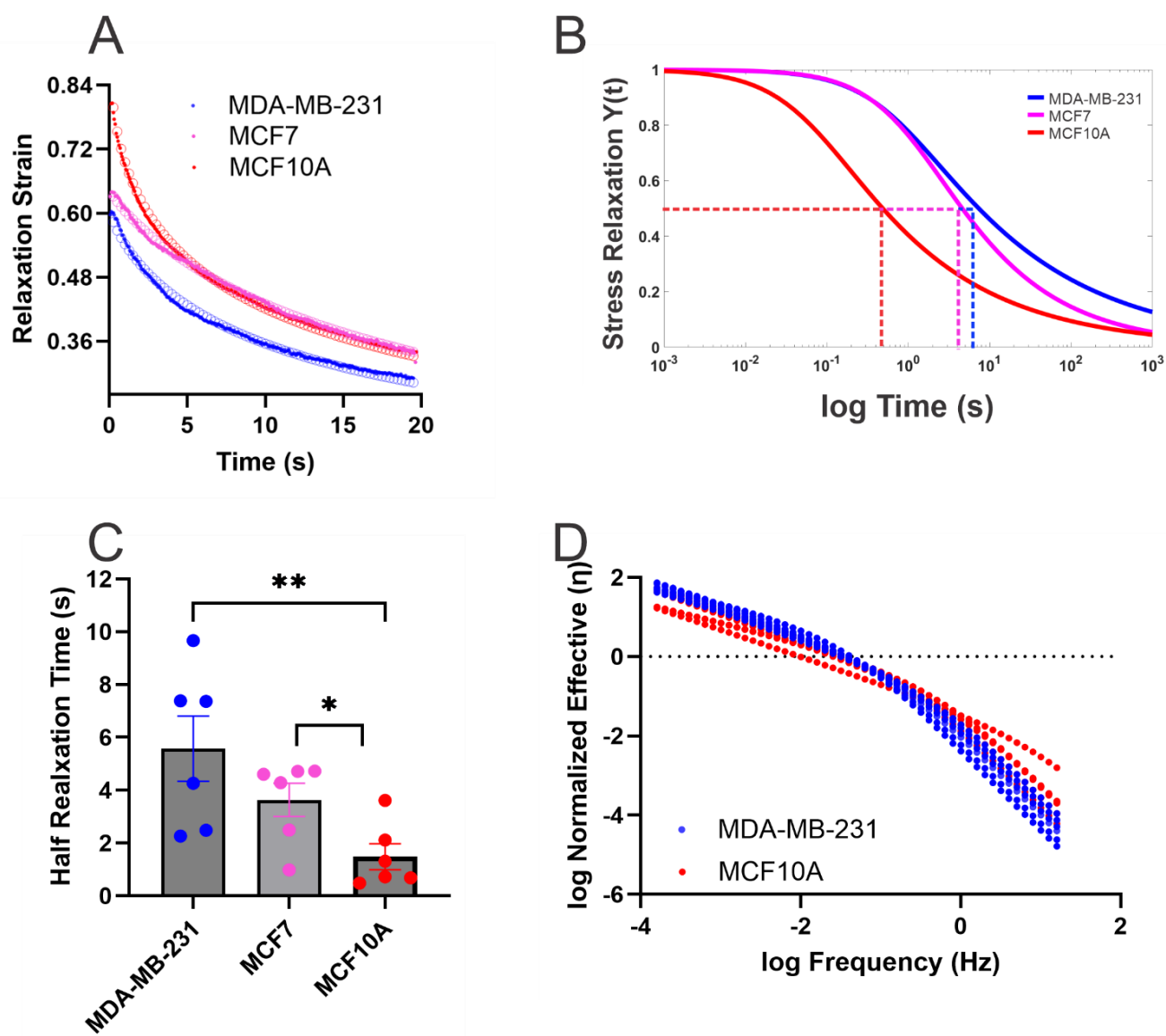

**Figure 3: Viscoelastic properties of tumor spheroids revealed by a modified power law model. (A)** The experimental strain response curve for spheroids of three different malignant levels. Circles are from experiments, and the lines are fits to the modified power law model. Here the strain is defined as $\Delta h/h$, where h is the vertical height of the spheroid, and is computed using the area change data shown in Fig. 2. **(B)** Stress relaxation curve computed using the strain response curve in (A) and the modified power-law model for three types of spheroids. Dashed lines indicate the location of half time when the stress decreases to half its original value. The fit data shown in 3 A and the stress relaxation curve in 3B is from a single spheroid for each individual cell line. **(C)** Half relaxation time of all three types of spheroids. The average value for MDA-MB-231 spheroids is 5.571 ± 1.34 s, MCF 7 is 3.636 ± 0.693 and MCF10A is 1.488 ± 0.488 s. **(D)** Computed effective viscosity of the spheroids over a range of frequency. The stars were obtained using a nonparametric t-test (Mann–Whitney test with ****: P < 0.0001, ***: P < 0.001, **: P < 0.01 and *: P < 0.05).

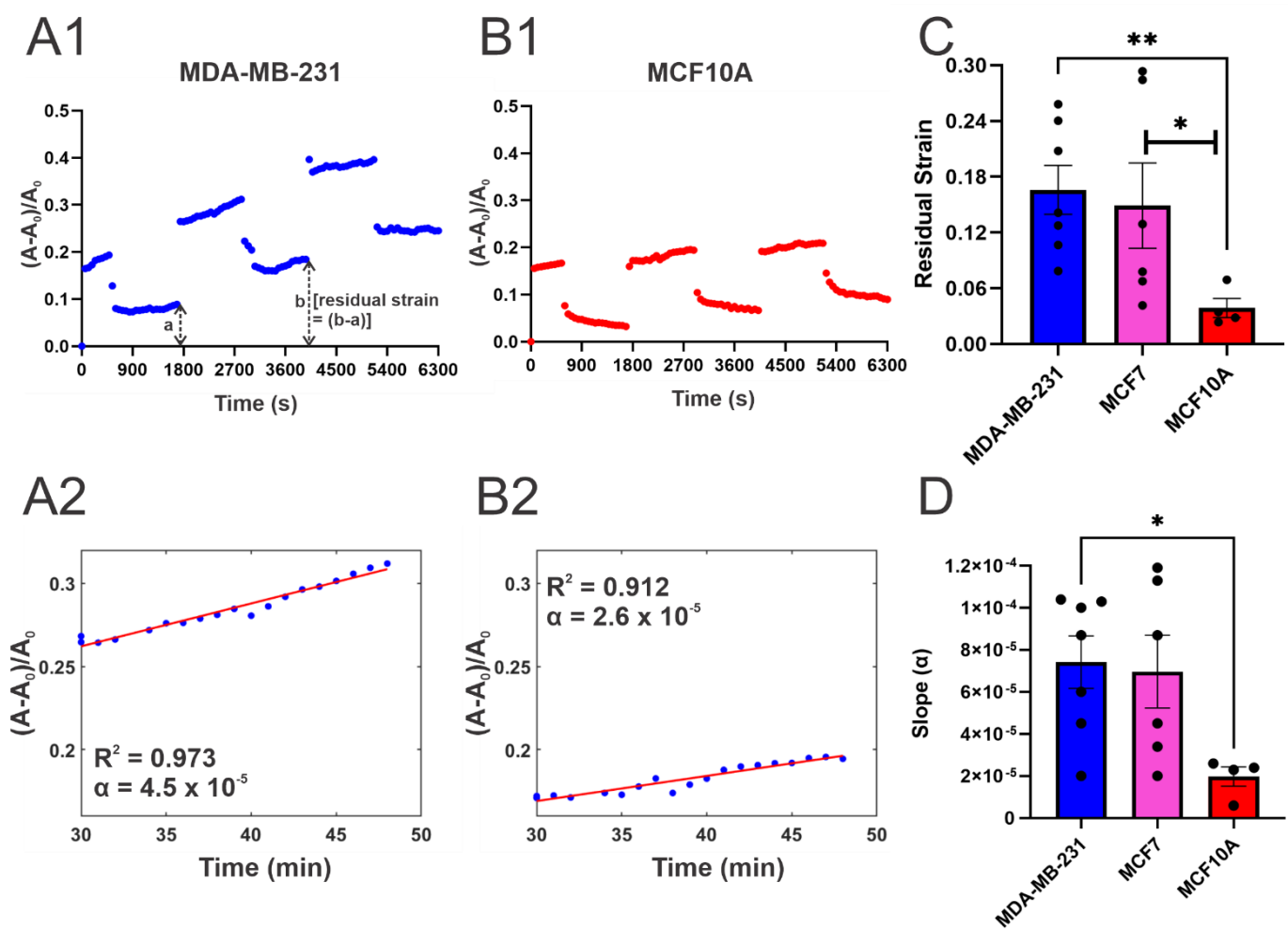

**Figure4: Plasticity of tumor spheroids revealed by the spheroid size response curves under long term compression. (**A1**)** Normalized spheroid area change under cyclic compression of MDA-MB-231 tumor spheroids. (A2) A linear fit to data in the compressed phase of the spheroid for MDA-MB-231 spheroid. (**B**1**)** Normalized spheroid area change under cyclic compression of MCF10A spheroids. (B2) A linear fit to data in the compressed phase of the spheroid for MCF10A spheroids. (C) Average residual strain calculated as the difference of strain recovery between cycle 2 and cycle 1 of cyclic compression (displayed graphically in A1). (**D**) The spheroid expansion rate during compressed phase for all different tumor spheroids. All the spheroids were embedded in 1.5 mg/ml collagen matrix and a maximum pressure is 14kPa square wave with a period of 40 minutes was applied. The stars were obtained using a nonparametric t-test (Mann–Whitney test with [****]: P < 0.0001, [***]: P < 0.001, [**]: P < 0.01 and [*]: P < 0.05).

# ARTICLE

## Author contributions

MP and MW created the research project. MP, YJS and MW designed the project. MP carried out the experiments. MP and KR performed data analysis. BZ, and CYH carried out the theoretical modelling work shown in the supplementary information. The manuscript was collaboratively written by MP, JES, and MW, with contributions from all authors.

## Conflicts of interest

There are no conflicts to declare.

## Data availability

Data sets generated during the study are available from the corresponding author on reasonable request.

## Acknowledgements

This work was supported by a grant from the National Institute of Health (Grant No. R01CA221346). The microfluidic device used in this work was developed at the Cornell NanoScale Facility, a member of the National Nanotechnology Coordinated Infrastructure (NNCI), which is supported by the National Science Foundation (Grant NNCI2025233). JES is the Betty and Sheldon Feinberg Senior Faculty Scholar in Cancer Research.

**Supplementary materials**

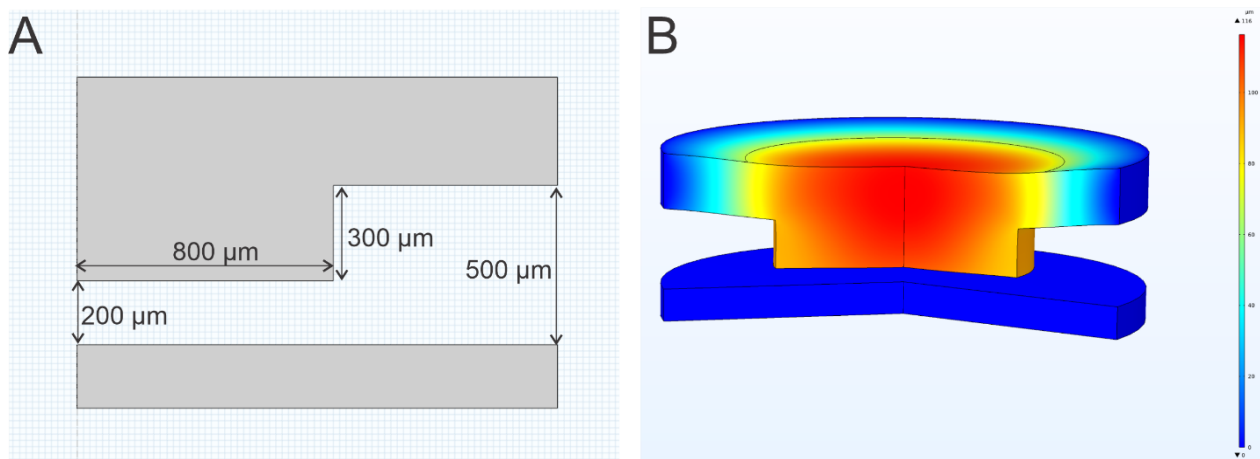

**Figure S1: (**A) Dimensions of a compression unit. (B) COMSOL simulation of vertical displacement of deformable piston layer upon application of 14kPa pressure.

**Theoretical models:**

1. **Modified Power law model to fit tumor spheroid compression data**

The relaxation function of the modified power-law model is given by[43],

$$Y(t) = E_\infty + \frac{E_0 - E_\infty}{\left(1 + \frac{t}{t_R}\right)^m} \tag{S1}$$

where $E_0 = Y(t = 0)$ is the instantaneous Young's modulus, $E_\infty = Y(t \to +\infty)$ is the relaxation Young's modulus or long-time modulus, $t_R$ and $m$ are two material fitting parameters. The modulus smoothly decreases from $E_0$ to $E_\infty$ with increasing time. Here, we define the ratio of instantaneous modulus and relaxation modulus as $\rho = E_0/E_\infty > 1$

An approximate expression for the corresponding creep function $C(t)$ associated with this $Y(t)$ is given in [44]

$$C(t) \approx \frac{Y(t)}{Y^2(t) + \left(\frac{\pi t}{2}\right)^2 \left[\frac{d\, Y(t)}{d\, t}\right]^2} \tag{S2}$$

Substituting Eq. (S3) into Eq. (S4),

$$C(t) \approx \frac{E_\infty + \dfrac{E_0 - E_\infty}{\left(1 + \frac{t}{t_R}\right)^m}}{\left[E_\infty + \dfrac{E_0 - E_\infty}{\left(1 + \frac{t}{t_R}\right)^m}\right]^2 + \left(\dfrac{m\pi t}{2t_R}\right)^2 \left[\dfrac{E_0 - E_\infty}{\left(1 + \frac{t}{t_R}\right)^{m+1}}\right]^2} \tag{S3}$$

We approximate the deformation of the spheroid during one loading-unloading cycle by a 1D uniaxial linear viscoelastic model where the uniaxial strain $\varepsilon$ is identified with $\Delta h/h$ . The strain history is:

a. At $t = 0$, a sudden compression strain $\varepsilon_0$ is applied. Note in the following discussion, we take the absolute values of all compression strain, so all strains values are positive.

b. For $0 < t < T$, strain is held constant. The system is under stress relaxation. The stress in this stage is given by, $\sigma(t) = Y(t)\varepsilon_0$. [44].

c. At $t = T$, the system is suddenly unloaded. Stress drops to zero, $\sigma(t > T^+) = 0$, and strain starts to decrease in this strain recovery stage.

The stress history can be expressed with the help of Heaviside step function $H(x) = \begin{cases} 1, x \geq 0 \\ 0, x < 0 \end{cases}$,

$$\sigma = Y(t)\varepsilon_0[H(t) - H(t - T)] \tag{S4}$$

The strain history from the Boltzmann superposition principle is,

$$\varepsilon(t) = \varepsilon_0 H(t) - \varepsilon_0 \int_0^t C(t - \tau) \frac{d[Y(\tau)H(\tau - T)]}{d\tau} d\tau \tag{S5}$$

For $0 < t < T$, the integral in Eq.(S5) is zero and $\varepsilon(0 < t < T) = \varepsilon_0$.

For $t > T$, the Eq.(S5) leads to,

$$\varepsilon(t > T) = \varepsilon_0 - \varepsilon_0 \left[ C(t - T)Y(T) + \int_{T^+}^t C(t - \tau) \frac{dY(\tau)}{d\tau} d\tau \right] \tag{S6a}$$

$$\overline{\varepsilon}(t > T) \triangleq \frac{\varepsilon(t > T)}{\varepsilon_0} = 1 - \frac{\left[1 + \dfrac{\rho - 1}{\left(1 + \dfrac{t - T}{t_R}\right)^m}\right]\left[1 + \dfrac{\rho - 1}{\left(1 + \dfrac{T}{t_R}\right)^m}\right]}{\left[1 + \dfrac{\rho - 1}{\left(1 + \dfrac{t - T}{t_R}\right)^m}\right]^2 + \left[\dfrac{m\pi(t - T)}{2t_R}\right]^2 \left[\dfrac{\rho - 1}{\left(1 + \dfrac{t - T}{t_R}\right)^{m+1}}\right]^2}$$

$$+ \frac{m}{t_R} \int_{T^+}^{t} \frac{\left[1 + \dfrac{\rho - 1}{\left(1 + \dfrac{t - \tau}{t_R}\right)^m}\right]\left[\dfrac{\rho - 1}{\left(1 + \dfrac{\tau}{t_R}\right)^{m+1}}\right]}{\left[1 + \dfrac{\rho - 1}{\left(1 + \dfrac{t - \tau}{t_R}\right)^m}\right]^2 + \left[\dfrac{m\pi(t - \tau)}{2t_R}\right]^2 \left[\dfrac{\rho - 1}{\left(1 + \dfrac{t - \tau}{t_R}\right)^{m+1}}\right]^2} \, d\tau \qquad \text{(S6b)}$$

If we take the limit $t \to T^+$, we have the normalized strain at the instant after the sudden unloading given by,

$$\overline{\varepsilon}(T^+) = 1 - \frac{\left[1 + \dfrac{\rho - 1}{\left(1 + \dfrac{T}{t_R}\right)^m}\right]}{\rho} = 1 - \frac{1}{\rho} - \frac{\rho - 1}{\rho}\frac{1}{\left(1 + \dfrac{T}{t_R}\right)^m} \qquad \text{(S7a)}$$

which means the sudden strain drop at unloading $t = T$ is,

$$\varepsilon_{\text{drop}} = |\overline{\varepsilon}(T^+) - \overline{\varepsilon}(T^-)| = \left[\frac{1}{\rho} + \frac{\rho - 1}{\rho}\frac{1}{\left(1 + \dfrac{T}{t_R}\right)^m}\right] \qquad \text{(S7b)}$$

This 1D theory explains the sudden strain drop and subsequent creep observed in our experiments. Experimental data fitting using Eq.(S6b) and Eq.(S7b) suggests $\rho \gg 1$, likely due to active rearrangements in the living tumor spheroid, consistent with a fluid-like long-time response ($E_\infty \to 0$).

Considering the time scale in our experiments, we can safely assume $E_\infty = 0, \rho \to +\infty$ in the above equations and take $t_R, m$ as the only fitting parameters. In the $\rho \to +\infty$ limit, the normalized strain becomes,

$$\overline{\varepsilon}(t > T) = 1 - \frac{\left(1 + \dfrac{t - T}{t_R}\right)^{m+2}}{\left(1 + \dfrac{t - T}{t_R}\right)^2 + \left[\dfrac{m\pi(t - T)}{2t_R}\right]^2}\left[\frac{1}{\left(1 + \dfrac{T}{t_R}\right)^m}\right]$$

$$+ \frac{m}{t_R} \int_{T^+}^{t} \frac{\left(1 + \dfrac{t - \tau}{t_R}\right)^{m+2}}{\left(1 + \dfrac{t - \tau}{t_R}\right)^2 + \left[\dfrac{m\pi(t - \tau)}{2t_R}\right]^2}\left[\frac{1}{\left(1 + \dfrac{\tau}{t_R}\right)^{m+1}}\right] d\tau \qquad \text{(S8a)}$$

$$\varepsilon_{\text{drop}} = |\overline{\varepsilon}(T^+) - \overline{\varepsilon}(T^-)| = \frac{1}{\left(1 + \dfrac{T}{t_R}\right)^m} \qquad \text{(S8b)}$$

The Eq. (S8a, b) are used to obtain $t_R, m$ from fitting the sudden drop of strain after unloading and the following gradual strain decreasing in the creeping stage.

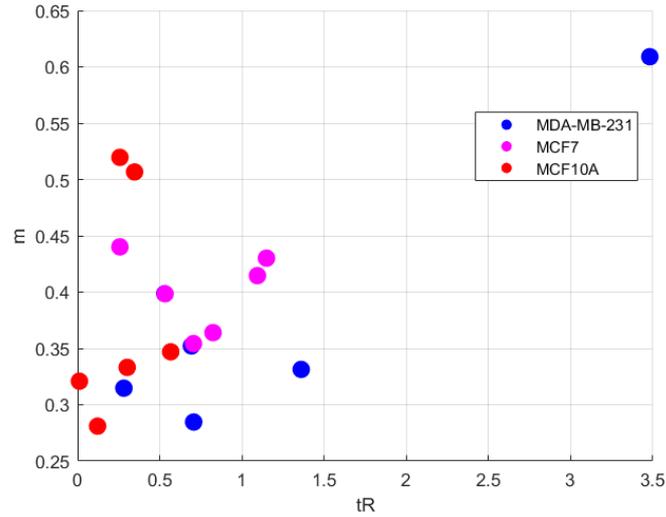

**Fig S2:** Fitted parameter values for $m$ and $t_R$, values for spheroids made of three different cell types, very malignant (MDA-MB-231), moderately malignant (MCF7) and non-tumorigenic (MCF10A) breast tumor cells. Each dot fits the data of response curve of one spheroid.

2. **Effective viscosity calculation**

In linear viscoelastic theory, the energy dissipated per cycle under sinusoidal strain $\varepsilon(t) = \varepsilon_0 \cos\omega t$ is given by,

$$W_D = \varepsilon_0^2 \omega E''(\omega) \int_0^{\frac{2\pi}{\omega}} \sin^2(\omega t)\, dt = \pi \varepsilon_0^2 \omega E''(\omega) \qquad (S9)$$

If we assume that at any fixed frequency $\omega$, the dissipation is due to a dashpot with viscosity $\eta(\omega)$, analogous to a Newtonian fluid,

$$\sigma = \eta(\omega)\dot{\varepsilon} \qquad (S10)$$

This leads to the energy dissipation per cycle,

$$W_D = \int_0^{\frac{2\pi}{\omega}} \sigma d\varepsilon = \int_0^{\frac{2\pi}{\omega}} \eta(\omega)\dot{\varepsilon} d\varepsilon = -\varepsilon_0 \omega \eta(\omega) \int_0^{\frac{2\pi}{\omega}} \sin(\omega t)\, d\varepsilon$$

$$= \varepsilon_0^2 \omega^2 \eta(\omega) \int_0^{\frac{2\pi}{\omega}} \sin^2(\omega t)\, dt = \pi \varepsilon_0^2 \omega^2 \eta(\omega) \qquad (S11)$$

Equating Eq.(S9) and Eq.(S11) gives,

$$\eta(\omega) = \frac{E''(\omega)}{\omega} \qquad (S12)$$

Thus, we can define an effective viscosity via Eq.(S14) using the loss modulus $E''(\omega)$. In some literature, the viscosity defined in Eq. (S14) is also called the dynamic viscosity [45].

From Eqn S1, with $E_\infty = 0$, the modified power-law relaxation function becomes,

$$Y(t) = \frac{E_0}{\left(1 + \frac{t}{t_R}\right)^m} \qquad (S13)$$

The complex modulus associated with this relaxation function is,

$$E^*(\omega) = i\omega \int_0^\infty Y(t)e^{-i\omega t}\mathrm{d}t = i\omega E_0 \int_0^\infty \left(1 + \frac{t}{t_R}\right)^{-m} e^{-i\omega t}\mathrm{d}t = E_0 e^c c^m \Gamma(1-m, c) \qquad (S14)$$

where $c = i\omega t_R$ and $\Gamma(s, c) = \int_c^\infty z^{s-1} e^{-z} dz$ is the upper incomplete gamma function.

An alternative equivalent expression using the Kummer's function $U(1, 1 - m, i\omega t_R) = \int_0^\infty \frac{e^{-i\omega t_R v}}{(1+v)^{m+1}} dv$ is,

$$E^*(\omega) = E_0 \left[ 1 - \frac{m}{t_R} \int_0^\infty \frac{e^{-i\omega t} dt}{\left(1 + \frac{t}{t_R}\right)^{m+1}} \right] = E_0[1 - mU(1, 1 - m, i\omega t_R)] \tag{S15}$$

Therefore, the effective viscosity $\eta(\omega)$ normalized by instantaneous modulus $E_0$ is given by,

$$\frac{\eta(\omega)}{E_0} = \frac{\mathrm{Im}[E^*(\omega)]}{E_0 \omega} = \frac{\mathrm{Im}[e^c c^m \Gamma(1 - m, c)]}{\omega}, c = i\omega t_R \tag{S16}$$

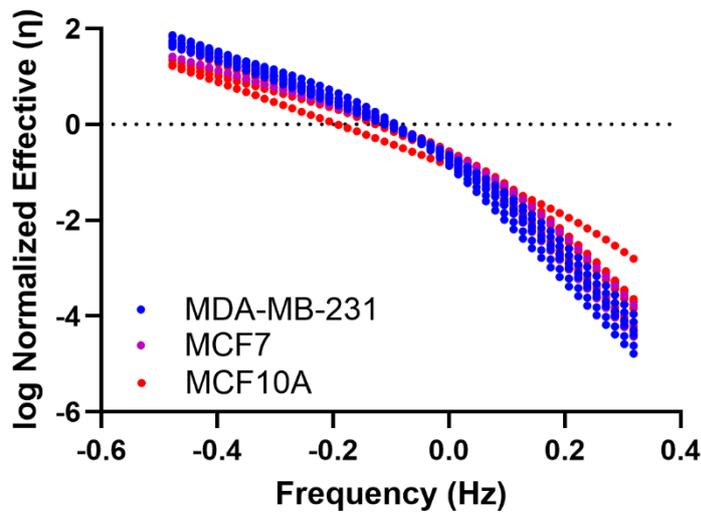

**Figure S3:** Normalized effective viscosity plot for MDA-MB-231, MCF7, and MCF10A breast cell spheroids.

3. **Conversion of measured spheroid diameter to height**

Compression of a spheroid under large deformation does not yield simple analytical results. Therefore, we employed the finite element method (FEM) to investigate the conversion from the change in area $\Delta A$ to the change in height $\Delta h$. The definition of each parameter is shown in Fig. S4.

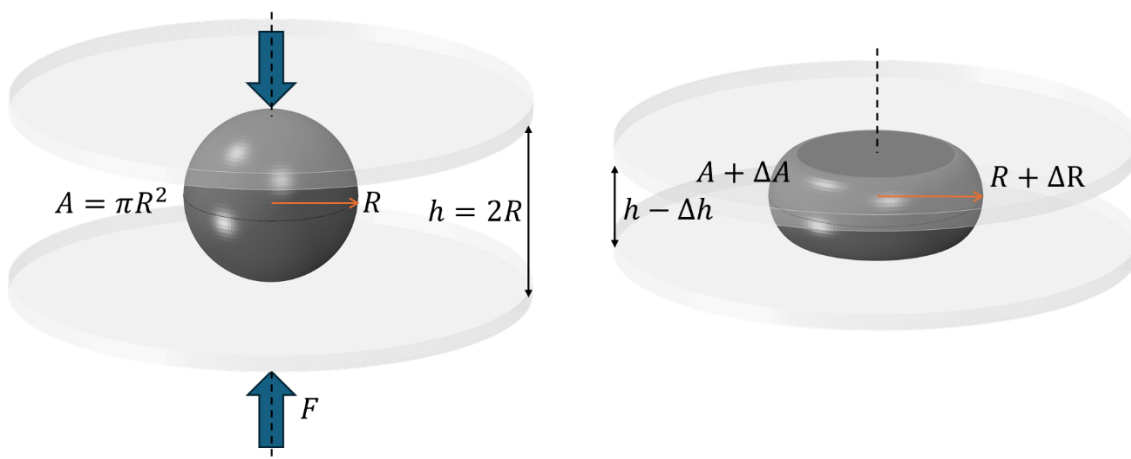

**Fig. S4** Illustration of a geometric relation of an initial stress-free (a, left) /compressed (b, right) sphere. Here, $F$ is the force exerted onto the sphere, $R$ is the radius of the sphere in the mid-z plane, $A$ is the area of the sphere in the mid-z plane, and $h$ is the height of the sphere. Left/right panel shows the sphere under initial stress-free/compressed state.

The simulation was conducted using Abaqus with axisymmetric hybrid elements (CAX4H). We assumed frictionless hard contact between the hyperelastic sphere and rigid plates and used an incompressible neo-Hookean material model for the spheroid.

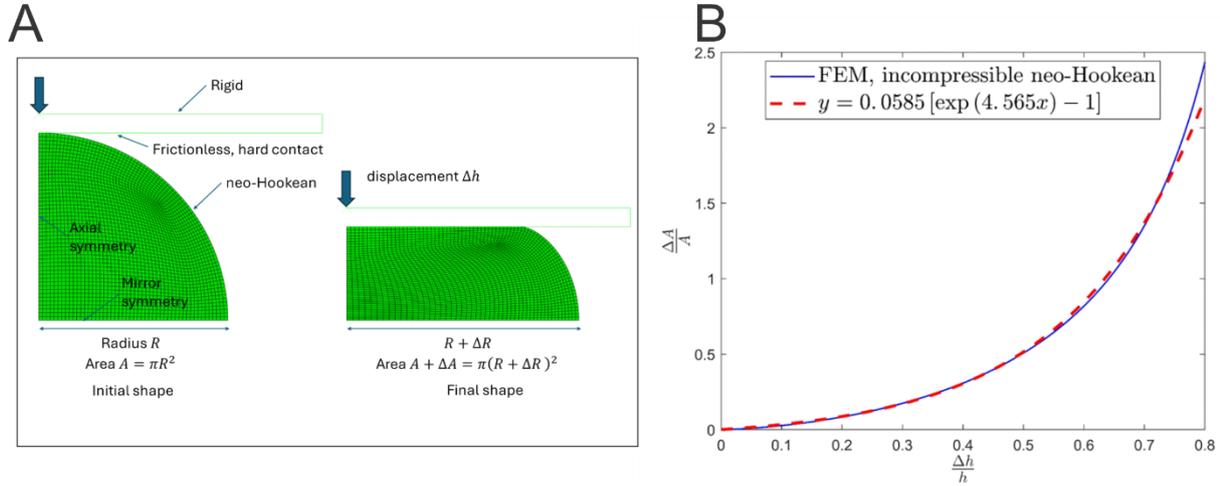

**Fig. S5 (A)** FEM computation setup. Here, material property is assumed to be neo-Hookean, and the volume is conserved. **(B)** Relation of area change versus height change computed from the FEM calculation. This relationship is independent of the radius of sphere and the elastic modulus, here $x = \Delta h/h$.

Under these conditions, a normalized master curve between $\Delta A/A$ and $\Delta h/h$ can be obtained from the simulation results, which does not depend on the spheroid radius $R$ and the Young's modulus $E$. Hence we conclude that this relation is purely geometry and is insensitive to the modulus.

We propose a simple exponential function to fit the FEM results:

$$\frac{\Delta A}{A} \cong 0.0585 \left[ e^{\left( 4.564 \frac{\Delta h}{h} \right)} - 1 \right] \tag{S17}$$

Although this relationship is derived in the compressed state, it is purely geometric in nature; thus, we use it as an approximation to analyze data in other states, such as during strain recovery stage after pressure release. The normalized strain $\bar{\varepsilon}(t)$ during strain recovery, $t > T^-$, is converted from the area change $\frac{\Delta A(t)}{A}$ using Eq.(S17). $T^-$ is the time just before unloading.

$$\bar{\varepsilon}(t) = \frac{\Delta h(t)}{\Delta h(T^-)} \cong \frac{\ln \left[ \frac{1}{0.0585} \frac{\Delta A(t)}{A} + 1 \right]}{\ln \left[ \frac{1}{0.0585} \frac{\Delta A(T^-)}{A} + 1 \right]} \cong \frac{\ln \left[ 17.1 \frac{\Delta A(t)}{A} + 1 \right]}{\ln \left[ 17.1 \frac{\Delta A(T^-)}{A} + 1 \right]} \tag{S18}$$

ArXiv

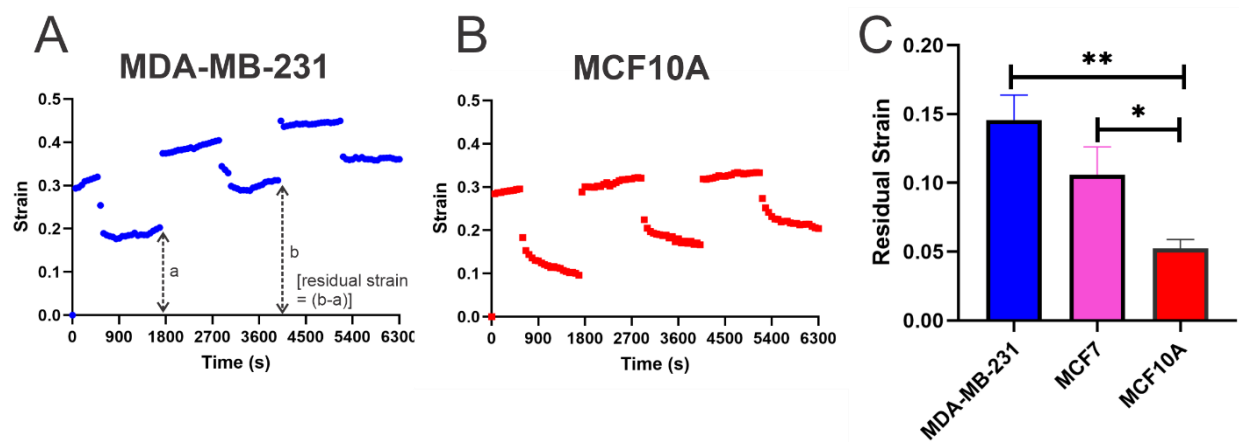

**Figure S6:** Residual strain calculated using the real strain values obtained via change in Area to change in height conversion discussed in S2.

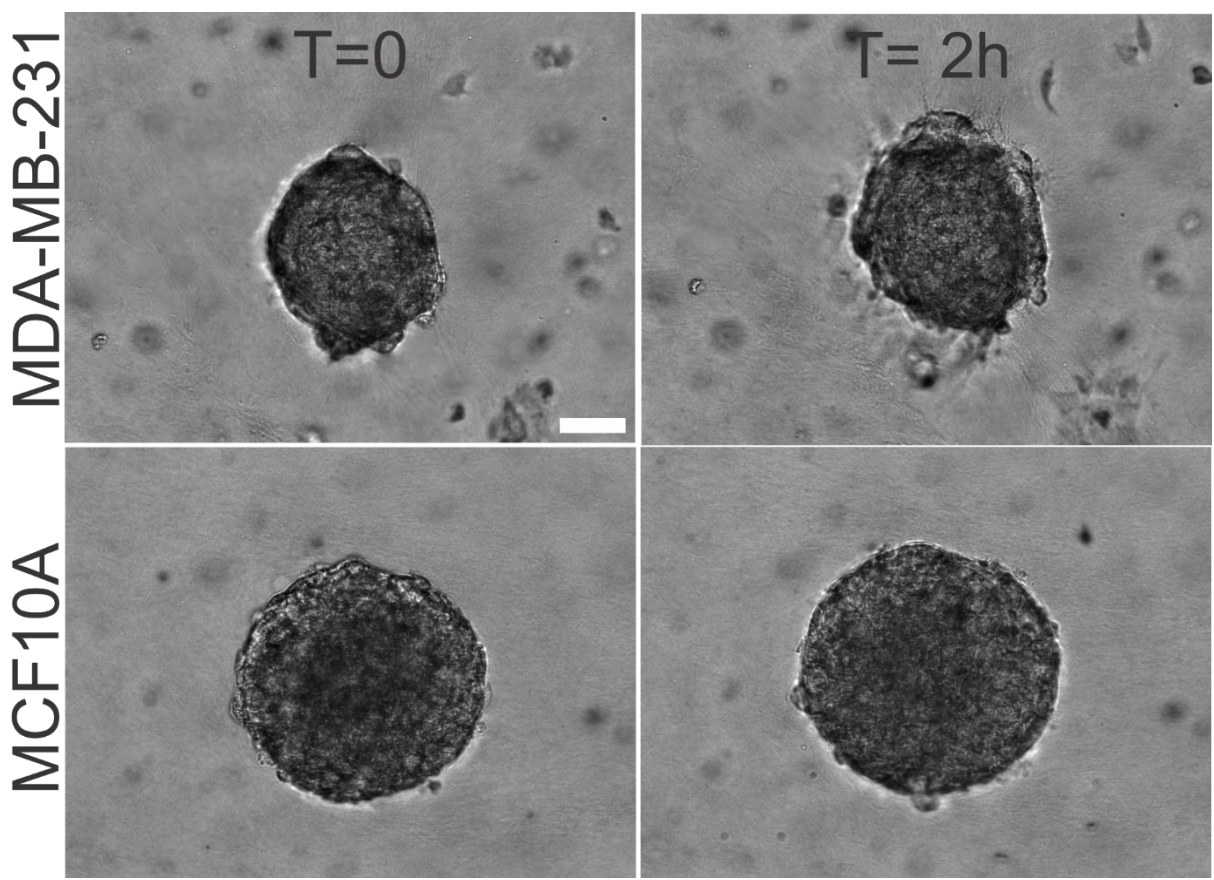

**Figure S7:** Malignant MDA-MB-231 spheroids begin to show protrusions indicating invasion at the end of three cyclic compressions. Malignant MDA-MB-231 and non-malignant MCF10A spheroids at the start and end of T = 40 min cyclic compression cycle. The scale bar is 50 μm.